\documentclass[twocolumn,showpacs,preprintnumbers,amsmath,amssymb,superscriptaddress, bibnotes]{revtex4}

\usepackage{graphicx}
\usepackage{dcolumn}
\usepackage{bm}
\usepackage{xspace}

\newcommand{\EF}{$E_\mathrm{F}$\xspace}
\newcommand{\Uf}{$\mathrm{U}~5f$\xspace}
\newcommand{\Cef}{$\mathrm{Ce}~4f$\xspace}
\newcommand{\orb}[2]{$\mathrm{ #1 } ~ #2 $\xspace}
\newcommand{\hn}[1]{$h\nu #1~\mathrm{eV}$\xspace}
\newcommand{\EB}[1]{$E_{\mathrm{B}} #1~\mathrm{eV}$\xspace}
\newcommand{\EBD}[2]{$E_{\mathrm{B}} #1$--$#2~\mathrm{eV}$\xspace}
\newcommand{\pnt}[1]{$\mathrm{#1}$\xspace}
\newcommand{\pntX}[2]{$\mathrm{#1}_{(#2)}$\xspace}
\newcommand{\Gm}{$\mathrm{\Gamma}$\xspace}
\newcommand{\lineX}[4]{$\mathrm{#1}_{(-#3,0,#4)}$--$\mathrm{#2}_{(0,0,#4)}$--$\mathrm{#1}_{(#3,0,#4)} $\xspace}
\newcommand{\lineXX}[2]{$\mathrm{#1}$--$\mathrm{#2} $\xspace}
\newcommand{\lineXXX}[3]{$\mathrm{#1}$--$\mathrm{#2}$--$\mathrm{#3} $\xspace}

\newcommand{\etal}{\textit{et al.}\xspace}

\newcommand{\Th}{$\mathrm{Th}$\xspace}

\newcommand{\URuSi}{$\mathrm{URu}_2\mathrm{Si}_2$\xspace}
\newcommand{\ThRuSi}{$\mathrm{ThRu}_2\mathrm{Si}_2$\xspace}
\newcommand{\CeRuSi}{$\mathrm{CeRu}_2\mathrm{Si}_2$\xspace}
\newcommand{\LaRuSi}{$\mathrm{LaRu}_2\mathrm{Si}_2$\xspace}

\begin{document}
\draft
\preprint{HEP/123-qed}

\title{Electronic structure of \ThRuSi studied by angle-resolved photoelectron spectroscopy: Elucidating the contribution of \Uf states in \URuSi}

\author{Shin-ichi~Fujimori}
\affiliation{Materials Sciences Research Center, Japan Atomic Energy Agency, Sayo, Hyogo 679-5148, Japan}

\author{Masaaki~Kobata}
\affiliation{Materials Sciences Research Center, Japan Atomic Energy Agency, Sayo, Hyogo 679-5148, Japan}

\author{Yukiharu~Takeda}
\affiliation{Materials Sciences Research Center, Japan Atomic Energy Agency, Sayo, Hyogo 679-5148, Japan}

\author{Tetsuo~Okane}
\affiliation{Materials Sciences Research Center, Japan Atomic Energy Agency, Sayo, Hyogo 679-5148, Japan}

\author{Yuji~Saitoh}
\affiliation{Materials Sciences Research Center, Japan Atomic Energy Agency, Sayo, Hyogo 679-5148, Japan}

\author{Atsushi~Fujimori}
\affiliation{Materials Sciences Research Center, Japan Atomic Energy Agency, Sayo, Hyogo 679-5148, Japan}
\affiliation{Department of Physics, University of Tokyo, Hongo, Tokyo 113-0033, Japan}

\author{Hiroshi~Yamagami}
\affiliation{Materials Sciences Research Center, Japan Atomic Energy Agency, Sayo, Hyogo 679-5148, Japan}
\affiliation{Department of Physics, Faculty of Science, Kyoto Sangyo University, Kyoto 603-8555, Japan}

\author{Yuji~Matsumoto}
\altaffiliation[Present address: ]{Department of Physics, University of Toyama, Toyama 930-8555, Japan}
\affiliation{Advanced Science Research Center, Japan Atomic Energy Agency, Tokai, Ibaraki 319-1195, Japan}

\author{Etsuji~Yamamoto}
\affiliation{Advanced Science Research Center, Japan Atomic Energy Agency, Tokai, Ibaraki 319-1195, Japan}

\author{Naoyuki~Tateiwa}
\affiliation{Advanced Science Research Center, Japan Atomic Energy Agency, Tokai, Ibaraki 319-1195, Japan}

\author{Yoshinori~Haga}
\affiliation{Advanced Science Research Center, Japan Atomic Energy Agency, Tokai, Ibaraki 319-1195, Japan}

\date{\today}

\begin{abstract}
The electronic structure of \ThRuSi was studied by angle-resolved photoelectron spectroscopy (ARPES) with incident photon energies of \hn{=655-745}.
Detailed band structure and the three-dimensional shapes of Fermi surfaces were derived experimentally, and their characteristic features were mostly explained by means of band structure calculations based on the density functional theory.
Comparison of  the experimental ARPES spectra of \ThRuSi with those of \URuSi shows that they have considerably different spectral profiles particularly in the energy range of $1~\mathrm{eV}$ from the Fermi level, suggesting that $\mathrm{U}~5f$ states are substantially hybridized in these bands.
The relationship between the ARPES spectra of \URuSi and \ThRuSi is very different from the one between the ARPES spectra of \CeRuSi and \LaRuSi, where the intrinsic difference in their spectra is limited only in the very vicinity of the Fermi energy.
The present result suggests that the \Uf electrons in \URuSi have strong hybridization with ligand states and have an essentially itinerant character.
\end{abstract}

\maketitle
\narrowtext
\section{INTRODUCTION}
Compounds with $AB_{2}X_{2}$ stoichiometry and the $\mathrm{ThCr_{2}Si_{2}}$-type crystal structure, abbreviated as ``122'' compounds, exhibit a rich variety of physical properties.
They consist of the alternating stacks of quasi-two-dimensional  $A$ and $B_{2}X_{2}$ layers, resulting in peculiar physical properties in this series of compounds \cite{Hoffmann}.
In recent years, iron-based superconductors with this crystal structure have been found \cite{BaKFe2As2}.
Furthermore, heavy fermion compounds such as $\mathrm{CeCu_{2}Si_{2}}$ \cite{CeCu2Si2}, $\mathrm{CeRu_{2}Si_{2}}$ \cite{CeRu2Si2_1}, and $\mathrm{YbRh_{2}Si_{2}}$ \cite{YbRh2Si2_NF} have the same crystal structure.
In particular, these heavy fermion compounds have been attracting much attentions due to their proximity to the quantum critical point \cite{SC122,Steglich122}.

In this class of materials, \URuSi holds a unique position owing to its enigmatic hidden order (HO) transition at $T_\mathrm{HO}=17.5~\mathrm{K}$ \cite{URu2Si2_Palstra}.
Its order parameter has not been identified in spite of extensive studies with various experimental and theoretical techniques for more than thirty years \cite{URu2Si2_review1,URu2Si2_review2}.
A number of angle-resolved photoelectron spectroscopy (ARPES) studies on \URuSi have been also performed, particularly after the discovery of the spectral change at the HO transition observed by Santander-Syro \etal \cite{URu2Si2_Andres1}.
Although a few common perceptions are shared among those studies, there remains unresolved issues \cite{URu2Si2_Tomasz_review,SF_review_JPCM, SF_review_JPSJ}.
In particular, the role of \Uf electrons in \URuSi remains under extensive debate.
In this regard, an understanding of the electronic structure of the non-$f$ electron analog \ThRuSi is essential to further comprehend the electronic structure of \URuSi and other compounds with the same crystal structure.

In the present paper, we report an ARPES study of \ThRuSi using soft x-ray synchrotron radiation (\hn{=665-735}).
The detailed band structure and Fermi surfaces of \ThRuSi were derived experimentally, and they are compared with the result of band structure calculation and the result of \URuSi.
It is shown that they were well explained by the band structure calculation based on the local density approximation.
Furthermore, the ARPES spectral profiles of \ThRuSi are considerably different from those of \URuSi, suggesting that \Uf states are strongly hybridized in \URuSi.

Denlinger \etal first reported an ARPES study of \ThRuSi, in which the band structure and Fermi surface map of \ThRuSi were obtained with incident photon energies of $h\nu=17$ and $31~\mathrm{eV}$ \cite{Denlinger_XRu2Si2}.
Although overall agreement between the experimental data and the band structure calculation was suggested, the detailed structure of the Fermi surface and band structure were not understood.
Meanwhile, dHvA experiments on \ThRuSi \cite{ThRu2Si2_dHvA1, ThRu2Si2_dHvA2} and its diluted alloy $\mathrm{U_{0.03}Th_{0.97}Ru_2 Si_2}$ \cite{UThRu2Si2_dHvA} have been reported, and most of the observed branches were well explained by band structure calculations.
However, the band structure and the location of each Fermi surface have been not understood yet.
They are particularly important information from the view point of comparing results between \ThRuSi and \URuSi.

Here, note that \ThRuSi is an archetypal compound of the $R^{4+}$ ($R =$ rare earth or actinide element) state with the $\mathrm{ThCr_{2}Si_{2}}$-type crystal structure \cite{ThRU2Si2_NMR} while a similar non-$f$ compound \LaRuSi has the $\mathrm{La^{3+}}$ state.
Therefore, a comparison of the electronic structure of this class of materials with \ThRuSi and \LaRuSi will provide good criteria to understand the valency of the rare-earth or the actinide element.
Another interesting aspect of studying \ThRuSi is that the electronic structures, including the topology of Fermi surface, of the materials in this class have certain similarities owing to their quasi-two-dimensional crystal structures.
Thus an understanding of the electronic structure of \ThRuSi is important for further understanding compounds with the same crystal structure.

\section{EXPERIMENTAL PROCEDURES}
\label{Exp}
Photoemission experiments were performed at the soft X-ray beamline BL23SU in SPring-8 \cite{BL23SU,BL23SU2}.
The overall energy resolution in angle-integrated photoemission (AIPES) experiments at \hn{=800} was about $140~\mathrm{meV}$, and that in ARPES experiments at \hn{=655-745} was $90-115~\mathrm{meV}$, depending on the photon energies.
The angular resolution of the ARPES experiments was about $\pm 0.15^\circ$, which corresponds to the momentum resolution of about $0.073~\mathrm{\AA^{-1}}$ at \hn{=735}.
High quality single crystals of \ThRuSi were grown by the Czochralski pulling method, as described in Ref.~\cite{ThRu2Si2_dHvA2}.
A clean sample surface was obtained by cleaving the samples in a UHV chamber.
The present ARPES experiment was performed for the cleaved surface perpendicular to the $c$ axis.
Therefore, the scans along the angular directions correspond to the scan within the $k_x$--$k_y$ plane while the scan along the photon energy corresponds to the scan along the $k_z$ direction.

Note that compounds with the $\mathrm{ThCr_{2}Si_{2}}$-type crystal structure have two types of cleaving surfaces, and each of them has very different photoemission spectral profiles \cite{Denlinger_XRu2Si2,Denis_YbRh2Si2}.
Although our photoemission experiments are less sensitive to surface electronic structures owing to highly enhanced kinetic energies of photoelectrons, two types of photoemission spectral profiles were certainly observed depending on the location of the cleaved surface.
One of them does not show clear periodic structures along the momentum direction and has lower photon energy dependences.
This suggests it is a signal from reconstructed surface electronic structures, which have an essentially two-dimensional nature with different periodicity of the Brillouin zone, as is expected from the bulk crystal structure.
The other set of spectra shows a clear periodic structure with the Brillouin zone of the bulk crystal structure, and has considerable photon energy dependencies.
Therefore, the latter set of spectra is considered as belonging to the bulk electronic structure, and the data presented in the paper pertain to that set.

The vacuum during the course of measurements was typically $<1 \times 10^{-8}~\mathrm{Pa}$, and the sample surfaces were stable for the duration of measurements ($1-2$ days) because no significant changes were observed in the ARPES spectra during the measurement period.
The positions of ARPES cuts were determined by assuming a free-electron final state where the momentum of electron is expressed as

\begin{eqnarray}
	k_\parallel  & = &  \frac{ \sqrt{2 m E_\mathrm{kin}}}{\hbar} \sin{\theta} - k_{\parallel \mathrm{photon}}, \nonumber\\
	k_\perp  & = & \sqrt{ \frac{2 m}{\hbar^2}(E_\mathrm{kin} \cos^2{\theta} + V_0) } - k_{\perp \mathrm{photon}}
	\label{kperp} 
\end{eqnarray}

We have taken the inner potential as $V_{0}=12~\mathrm{eV}$, which is common value of uranium based compounds \cite{SF_review_JPSJ}.
Note that the present ARPES experiment is insensitive to the value of $V_{0}$ since $E_\mathrm{kin}$ is more than one order larger than the inner potential.
For ARPES spectra, background contributions from elastically scattered photoelectrons due to surface disorder or phonons were subtracted by assuming momentum-independent spectra.
The details of the procedure are described in Ref.~\cite{UGe2_UCoGe_ARPES}. 

\section{RESULTS AND DISCUSSION}
\subsection{Angle-integrated photoemission spectra}
\begin{figure}
	\includegraphics[scale=0.5]{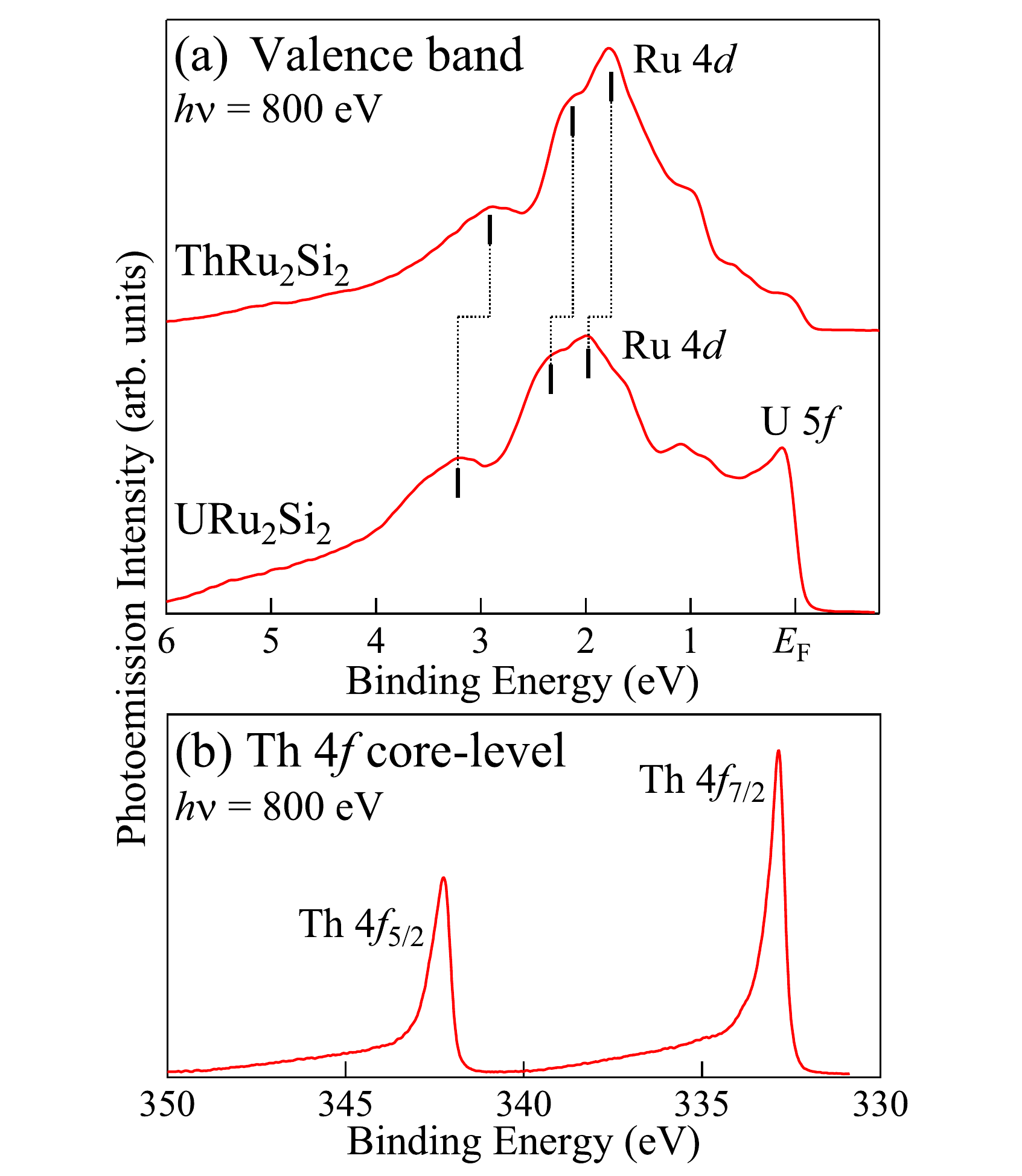}
	\caption{(Online color)
	Angle-integrated photoemission spectra of \ThRuSi measured at \hn{=800}.
	Data of \URuSi replotted from Ref.~\cite{Ucore}.
	(a) Valence-band spectra of \ThRuSi and \URuSi.
	The locations of pronounced peaks in the \orb{Ru}{4d} bands are indicated.
	(b) \orb{Th}{4f} core-level spectrum of \ThRuSi.
	}
\label{AIPES}
\end{figure}
Figure~\ref{AIPES} (a) shows the valence-band angle-integrated photoemission (AIPES) spectra of \ThRuSi and \URuSi.
The spectrum of \ThRuSi has a complex peak structure.
Pronounced peaks located at \EB{\gtrsim 1} originate mainly from \orb{Ru}{4d} states, which do not contribute majorly to the \EF.
The AIPES valence-band spectrum of \URuSi measured at \hn{=800} \cite{Ucore} is also shown in the bottom of Fig.~\ref{AIPES} (a).
The spectrum has a very similar spectral profile to that of \ThRuSi, except for the state in the vicinity of \EF.
The spectrum of \URuSi contains a sharp peak structure just below \EF, but the structure is absent in the spectrum of \ThRuSi.
This suggests that the specific state in the spectrum of \URuSi is contributed by \Uf states.
Interestingly, the locations of pronounced peaks in the \orb{Ru}{4d} bands of \URuSi are located at deeper binding energies by $\mathrm{0.2-0.3~eV}$ than those in \ThRuSi.
This is in contrast with the case of \CeRuSi and \LaRuSi where the positions of the \orb{Ru}{4d} bands are essentially identical \cite{Denlinger_XRu2Si2}.
This suggests that \Uf electrons contribute substantially to the valence band as itinerant electrons in \URuSi, while \Cef electrons behave almost as core electrons in \CeRuSi, and they do not majorly influence its valence-band structure.

Figure~\ref{AIPES} (b) shows the \orb{Th}{4f} core-level spectrum of \ThRuSi.
The spectrum has a relatively simple peak structure with an asymmetric shape having a long tail toward high binding energies.
The \orb{Th}{4f} core-level spectra of metallic \Th compounds are often accompanied by a satellite structure on the high-binding-energy side of $1.4-3.1~\mathrm{eV}$, and their spectral shapes have been analyzed based on the single impurity Anderson model (SIAM) \cite{Thcore_SIAM, Thcore_Sarma}.
The calculations suggest that the intensity of the satellite increases as the DOS at \EF increases.
Therefore, absence of the satellite structure in the core-level spectrum of \ThRuSi is consistent with the almost occupied nature of \orb{Ru}{4d} states in this compound.
The overall shape of the spectrum is very similar to that of \Th metal \cite{Baer_Th4f}.
This suggests that \Th is in the $\mathrm{Th^{4+}}$ state although there might finite contributions from \orb{Th}{5f} states in their valence bands \cite{Th_metal}.

\subsection{Band structure calculation}
\begin{figure}
	\includegraphics[scale=0.5]{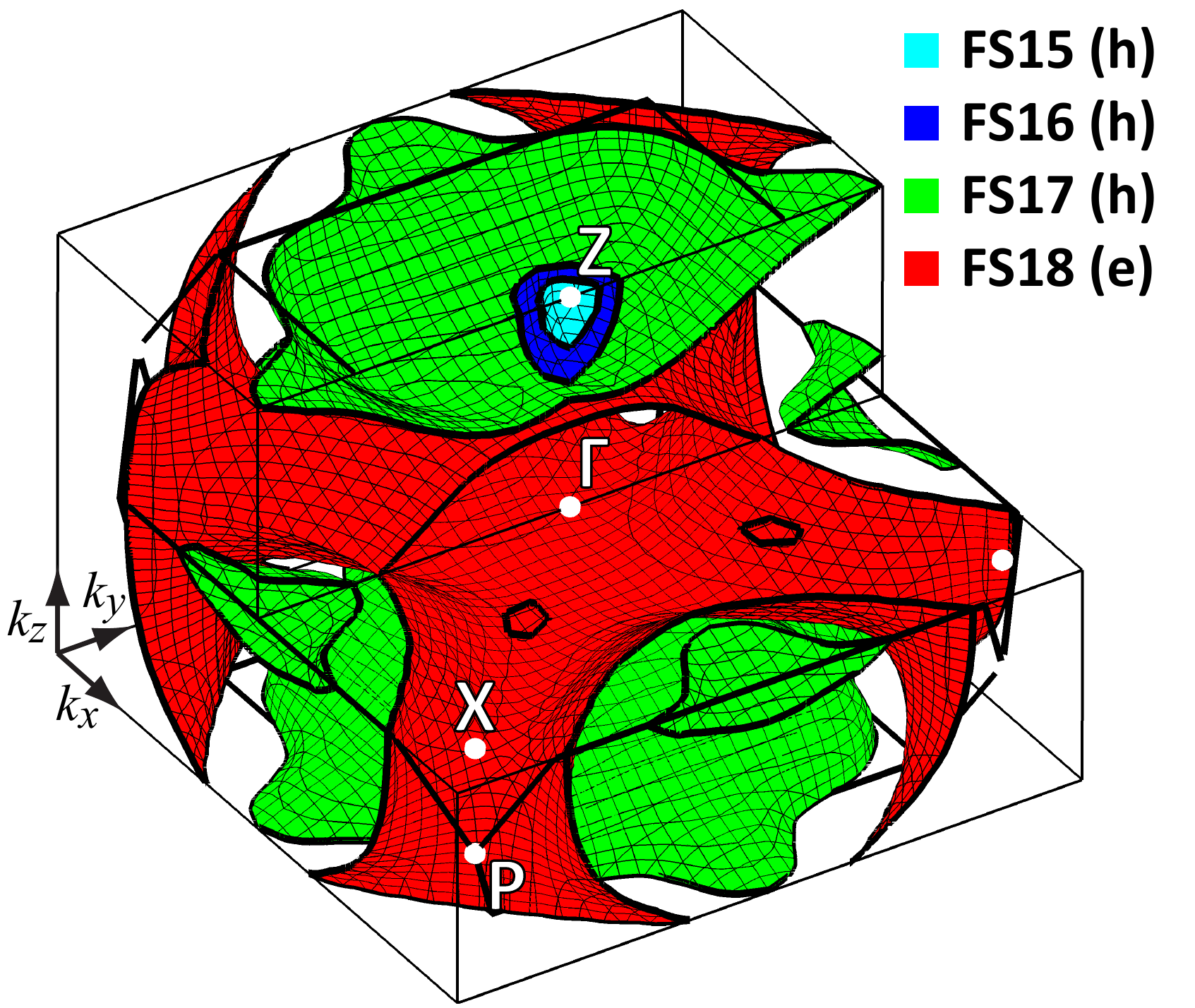}
	\caption{(Online color)
	Three-dimensional shape of calculated Fermi surface of \ThRuSi.
	Four bands form Fermi surfaces in the calculation.
	Bands 15 and 16 form small concentric hole-pockets centered at the \pnt{Z} point.
	Band 17 forms a closed hole-type Fermi surface centered at the \pnt{Z} point, which is referred to as the ``pillow.''
	Band 18 forms a multiply connected electron-type Fermi surface, which is referred to as  ``jungle-gym'' and a small hole-pocket almost in the middle of the \lineXX{\Gamma}{X} high-symmetry line.
	The total volume of the hole-type and the electron-type Fermi surfaces are equal because \ThRuSi is a compensated metal.
	}
\label{3DFS}
\end{figure}
\begin{figure*}[ht]
	\includegraphics[scale=0.5]{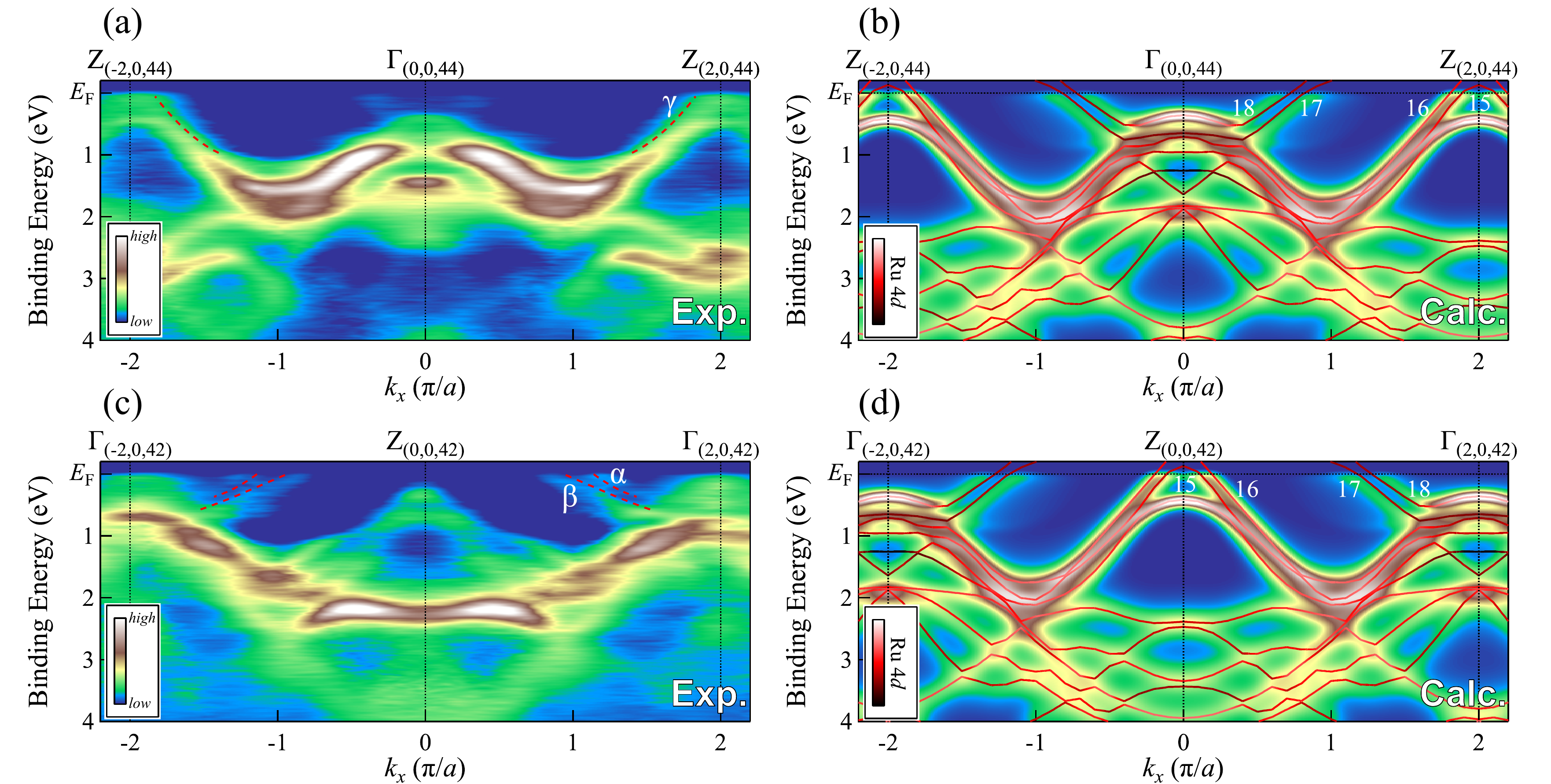}
	\caption{(Online color)
	Band structures of \ThRuSi obtained by photon energy scan of ARPES measurement, together with corresponding results of band structure calculation.
	(a) ARPES spectra measured along the \lineX{Z}{\Gamma}{2}{44} high-symmetry line.
	The denotation of the point represents the location of the momentum space in the units of $(\pi/a)$ for $k_x$ and $k_y$ directions and $(\pi/c)$ for $k_z$ direction, respectively (See Fig.~\ref{ARPES_hn_FS} (a)).	
	Approximate location of the band forming Fermi surface is indicated by dashed curves.
	(b) The corresponding calculated band structure and the simulation of ARPES spectra.
	(c) ARPES spectra measured along the \lineX{\Gamma}{Z}{2}{42} high-symmetry line.
	Approximate location of the bands forming the Fermi surface is indicated by the dashed curves.
	(d) Corresponding  calculated band structure and the simulated ARPES spectra.
	}
\label{ARPES_hn_band}
\end{figure*}

Before showing the experimental ARPES spectra, the result of band structure calculation in the present study and its relationship to previous studies are summarized.
Figure~\ref{3DFS} shows the three-dimensional shape of the calculated Fermi surface.
Four bands form Fermi surfaces in the band structure calculation.
Bands 15 and 16 form small hole-pockets at the \pnt{Z} point.
Band 17 forms a large closed Fermi surface centered at the \pnt{Z} point, which is referred to as ``pillow.''
Its volume is much larger than those of the surfaces formed by bands 15 and 16.
Band 18 forms a multiply connected and electron-type Fermi surface, referred to as ``jungle-gym.''
The total volume of the hole-type and electron-type Fermi surfaces are equal because \ThRuSi is a compensated metal.

Here, we state the relationship between the present Dirac-type relativistic LAPW calculation with those in Refs. \cite{Denlinger_XRu2Si2} and \cite{ThRu2Si2_dHvA2}.
The overall topology of the Fermi surface is almost identical to the one in those calculations, but there are a few minor differences:
The topology of the calculated Fermi surfaces in Ref.~\cite{Denlinger_XRu2Si2} is identical to the one in the present study, but their sizes and shapes are slightly different.
For example, the size of the small Fermi surface located at the center of the \lineXX{\Gamma}{X} direction is larger than that derived from the present calculation.
This minor difference might be due to improvements in the numerical calculations resulting from significant advances in computational resources, which have enabled us to use a much smaller mesh size.
Very similar topology of the calculated Fermi surface was reported in Ref.~\cite{ThRu2Si2_dHvA2}.
One of the major differences between this calculation and the present one is the absence of the electron-type Fermi surface 19 and the existence of the hole-type Fermi surface 15 in the present calculation.
These differences originate from the different treatments of lattice parameters and muffin-tin radii in these two sets of calculations.
Generally, first-principles band structure calculations based on the density-functional theory include some cutoff parameters which are needed to expand basis functions and potentials, and it is empirically known that energy band dispersions of ``122''-type compounds are sensitive to these parameters.
In the present calculation, the lattice constants and the muffin-tin radii of potential parameters are optimized by minimizing the total energy, whereas the experimental parameters were used in Ref.~\cite{ThRu2Si2_dHvA2}.
Note that the $z$ parameter, which is the $z$ position of Si atom at 4(e) site, was fixed in this procedure. 
By this procedure, agreement between the experimental and the calculated band structure improved considerably.
Therefore, we use the result of this calculation.
As we have learned, there are a few inherent issues in band structure calculations, although they have been considered {\it ab initio} calculations.
This matter is outside the scope of the present paper, and their details pertaining to it will be discussed elsewhere.

Interestingly, the overall shape of the Fermi surfaces is very similar to those of the Fermi surface of other rare-earth compounds with the same crystal structure.
For example, the renormalized band calculation of $\mathrm{YbRh_{2}Si_{2}}$ \cite{YbRh2Si2_FS} predicts that there exist similar ``pillow'' (``donut'' in Ref.~\cite{YbRh2Si2_FS}) and ``jungle-gym'' Fermi surfaces.
Therefore, they seem to be common features of compounds with the same crystal structure, which might be an origin of the unusual properties of the series of compounds.

In the present study, we have simulated the ARPES spectra based on the calculation.
In the simulation, the following effects were taken into account: (i) broadening along the $k_\perp$ direction due to finite escape depth of photoelectrons, (ii) lifetime broadening of the photohole, (iii) photoemission cross sections of orbitals, and (iv) energy resolution and angular resolution of the electron analyzer.
The broadening along $k_\perp$ direction is taken to be the inverse of the expected escape depth of photoelectrons in the present experiment ($15~\mathrm{\AA}$).
Life time broadening is assumed to have a linear dependence on $(E-E_{\rm F})$ on a wide energy scale as has been observed in Ni metal\cite{Ni_ARPES}.
We have assumed that it is zero at $E_{\rm F}$ and 0.5 eV at $E_{\rm B}=5$~eV.
The experimental energy and momentum resolutions were assumed to be experimental values as described in Sec.~\ref{Exp}.
The details are described in Ref.~\cite{UN_ARPES}.

\begin{figure}
	\includegraphics[scale=0.5]{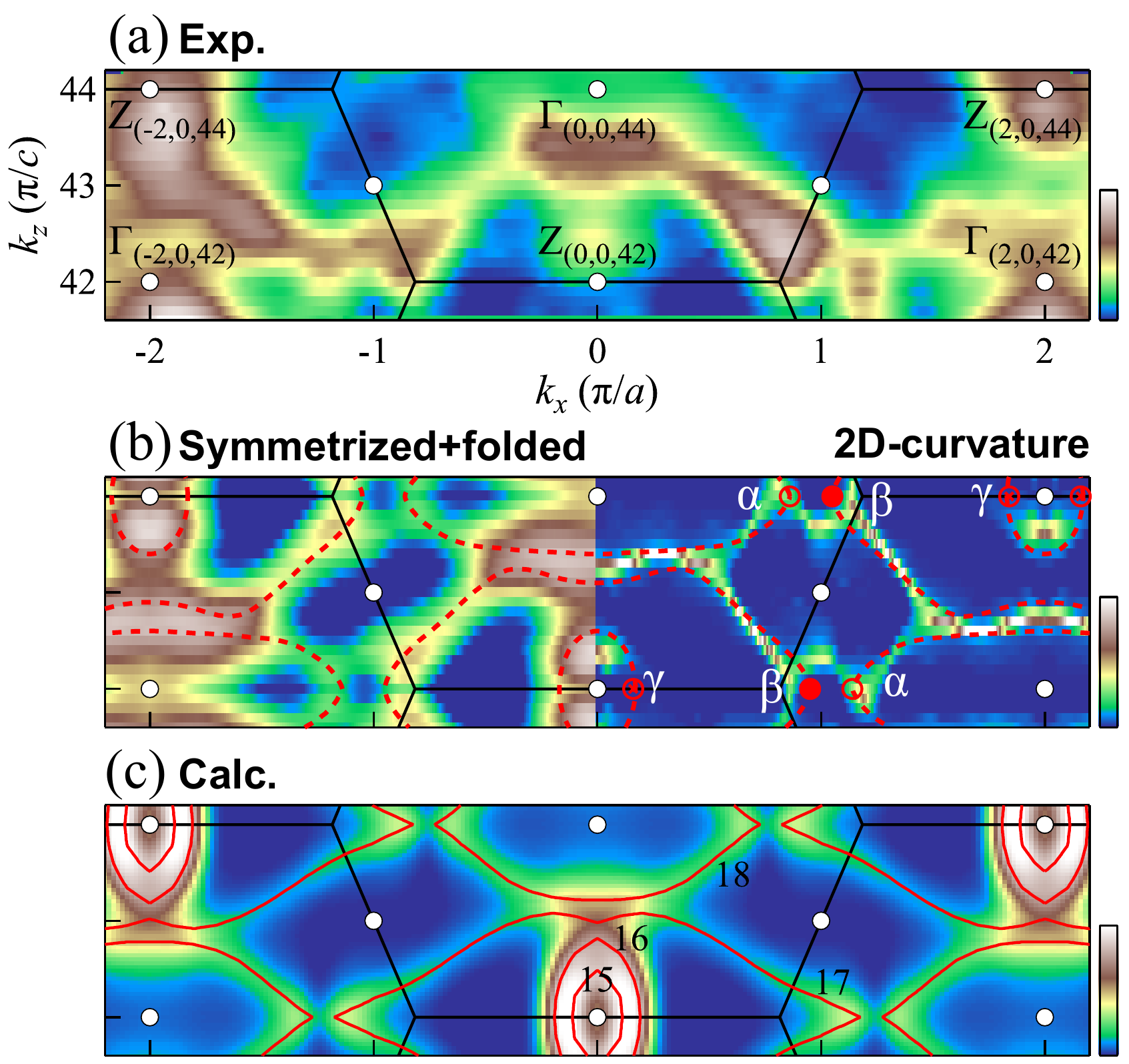}
	\caption{(Online color)
	Fermi surfaces of \ThRuSi obtained by photon energy scan of ARPES measurement, together with corresponding results of band structure calculation.
	(a) Fermi surfaces of \ThRuSi obtained by integrating ARPES spectra measured at $h\nu =655$--$745~\mathrm{eV}$ over $\mathrm{100~meV}$ at \EF.
	(b) Symmetrized and folded Fermi surface map within the $k_x$--$k_z$ plane (left panel) and obtained by the 2D curvature method \cite{curvature} (right panel).
	The locations of $\alpha$, $\beta$, and $\gamma$ bands derived from the band structures shown in Fig.~\ref{ARPES_hn_band} (a) and (c) are also indicated by filled circles.
	Approximate locations of Fermi surfaces derived from the Fermi surface map and the 2D-curvature analysis are indicated by dashed curves.
	(c) Calculated Fermi surfaces and the simulated Fermi surface map within the $k_x$--$k_z$ plane.
	}
\label{ARPES_hn_FS}
\end{figure}
\subsection{Electronic structure within $k_x$--$k_z$ plane}
We first show the ARPES spectra measured along $k_x$--$k_z$ plane, which correspond to scans along the detection angle of photoelectrons and incident photon energy.
Figure~\ref{ARPES_hn_band} (a) shows the ARPES spectra measured along the $\mathrm{Z}_{(-2,0,44)}$--$\mathrm{\Gamma}_{(0,0,44)}$--$\mathrm{Z}_{(2,0,44)}$ high-symmetry line.
Here, the denotation of the point represents the location of the momentum space in the units of $(\pi/a)$ for $k_x$ and $k_y$ directions and $(\pi/c)$ for $k_z$ direction, respectively (See Fig.~\ref{ARPES_hn_FS} (a)).
Note that the spectra correspond to the angular scan with \hn{\sim 735}, but contain spectra measured at its vicinity photon energies to exactly trace the high symmetry line according to the Eq.~\ref{kperp}.
Clear energy dispersions can be recognized.
On the higher binding energy side ($E_{\mathrm{B}} \gtrsim 1~\mathrm{eV}$), there exist weakly dispersive features with enhanced intensities, which are attributed to bands with the dominant contribution from \orb{Ru}{4d} states.
Moreover, there exist rapidly dispersing features in the vicinity of the Fermi energy, and some of them seem to form Fermi surfaces.
The hole-shaped dispersion centered at \pntX{Z}{2,0,44} and designated as $\gamma$ has almost linear energy dispersion toward \EF, suggesting that it forms hole-like Fermi surfaces at the \pnt{Z} point.
In addition, an electron-pocket-like feature with very weak intensities is observed at the \pntX{\Gamma}{0,0,44} point.
Its structure in the vicinity of \EF is particularly vague, and its nature is not well resolved.
There is a similar vague feature in the vicinity of \EF in the ARPES spectra measured along the same high-symmetry line at \hn{=17} \cite{Denlinger_XRu2Si2}, and its details were not well resolved.
One possibility is that this is the surface-related electronic structure because its intensity is highly enhanced in their spectra, which should have enhanced contributions from surface electronic structures.

Figure~\ref{ARPES_hn_band} (b) shows the calculated band structure and the simulation of ARPES spectra based on the band structure calculation.
The solid lines represent the band dispersions, and the intensity map shows the result of the simulation.
The color coding of the bands is the projection of the contributions from the \orb{Ru}{4d} states.
Very complicated band dispersions are expected in this energy region.
There are cosine-shaped dispersions with enhanced intensities in the binding energy of \EBD{=0.5}{1.5}, which have large contribution from the \orb{Ru}{4d} states.
Although their energy dispersions are somewhat larger than those in the experimental ARPES spectra, their essential structure is very similar to that observed in the experiment.
In the vicinity of the \pnt{Z} point, there exist hole pockets formed by bands 15 and 16.
These two bands are so close in momentum that they cannot be observed separately in the simulated ARPES spectra.
Therefore, although the feature is very similar to the experimental spectra, the number of Fermi surfaces in the vicinity of the \pnt{Z} point cannot be distinguished experimentally.
Meanwhile, the features corresponding to bands 17 and 18 are not observed clearly in the experimental ARPES spectra.

Figure~\ref{ARPES_hn_band} (c) shows the ARPES spectra measured along the $\mathrm{\Gamma}_{(-2,0,42)}$--$\mathrm{Z}_{(0,0,42)}$--$\mathrm{\Gamma}_{(2,0,42)}$ high-symmetry line.
The spectra correspond to the angular scan with \hn{\sim 665}.
The spectra are essentially very similar to those measured along  the $\mathrm{Z}_{(-2,0,44)}$--$\mathrm{\Gamma}_{(0,0,44)}$--$\mathrm{Z}_{(2,0,44)}$ high-symmetry line shown in Fig.~\ref{ARPES_hn_band} (a), but there are some distinct differences between them.
One of the major differences is the existence of clear dispersive features in the middle of the $\mathrm{\Gamma}_{(0,0,42)}$--$\mathrm{Z}_{(2,0,42)}$ high-symmetry line in the vicinity of \EF.
There are clear hole-like dispersions around $\mathrm{Z}_{(0,0,42)}$ point, which are designated as $\alpha$ and $\beta$.
They did not appear in the ARPES spectra measured along the $\mathrm{\Gamma}_{(-2,0,44)}$--$\mathrm{Z}_{(0,0,44)}$--$\mathrm{\Gamma}_{(2,0,44)}$ high-symmetry line.
Similar phenomena have been also observed in the ARPES spectra of other uranium compounds \cite{UN_ARPES,URhGe_ARPES}, and they are considered to have originated from the photoemission structure factor (PSF) effect \cite{PSF}.
This effect seems to be particularly significant in compounds with this type of crystal structure \cite{EuAl4}.
Therefore, it is quite important to observe multiple locations in the Brillouin zone with different photon energies and detection angles to deduce the entire electronic structure.

Figure~\ref{ARPES_hn_band} (d) shows the corresponding calculated band structure and the simulated ARPES spectra based on the band structure calculation.
There exist bands 17 and 18, which form a large hole-like Fermi surface around the \pnt{Z} point and  a large electron-like Fermi surface around the \Gm point; these structures correspond well to band $\beta$ and $\alpha$ in the experimental ARPES spectra.
Accordingly, the band structure along the \lineXXX{Z}{\Gamma}{Z} high-symmetry line is essentially explained by the band structure calculation although the number of small Fermi surface around \pnt{Z} point could not be resolved in the present experiment.

\begin{figure*}[ht]
	\includegraphics[scale=0.5]{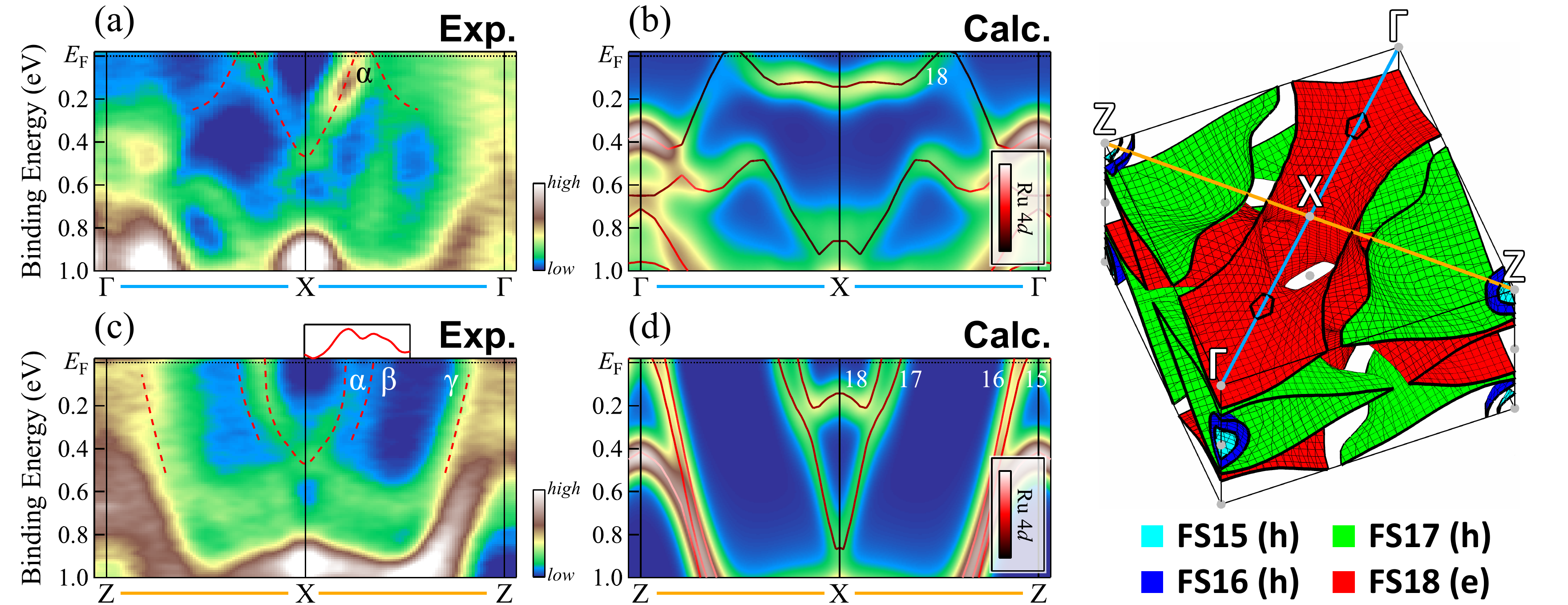}
	\caption{(Online color)
	Band structures of \ThRuSi obtained by angular scan of ARPES spectra measured at \hn{=735}, together with the corresponding results of the band structure calculation.
	(a) ARPES spectra measured along the \lineXXX{\Gamma}{X}{\Gamma} high-symmetry line.
	Approximate band location is indicated by the dashed curve $\alpha$, which forms a small hole-pocket along the path from the \pnt{X} point to the \Gm point.
	(b) Corresponding calculated band structure and the simulated of ARPES spectra.
	(c) ARPES spectra measured along the \lineXXX{Z}{X}{Z} high-symmetry line.
	Approximate locations of the two bands are designated by the dashed curves $\alpha$, $\beta$, and $\gamma$.
	The upper panel shows the MDC spectrum at \EF.
	(d) The corresponding  calculated band structure and the simulated ARPES spectra.
	Three-dimensional shape of calculated Fermi surface within the \lineXXX{\Gamma}{Z}{X} plane is also shown. 
	}
\label{ARPES_kxky_band}
\end{figure*}
Next, we discuss the shapes of the Fermi surfaces obtained by these measurements.
Figure~\ref{ARPES_hn_FS} (a) shows the Fermi surface map along the $k_x$--$k_z$ plane obtained by integrating ARPES spectra measured at $h\nu =655$--$745~\mathrm{eV}$ with $5~\mathrm{eV}$ step over $100~\mathrm{meV}$ at \EF.
In this Fermi surface map, the momenta parallel and perpendicular to the surface were calculated based on Eq.~\ref{kperp}.
The Fermi surface map has complex features.
Note that locations with the same symmetry in the Brillouin zone but with different values of $k_x$ and $k_z$ have different profiles.
For example, there exists the feature with an enhanced intensity centered at the \pntX{Z}{0,0,42} point ranging over $k_x = -1$ -- $1 (\pi/a)$ and $k_z =42$ -- $43.5 (\pi/c)$, but the corresponding features are very vague around the \pntX{Z}{\pm 2,0,44} points.
Another example is a vertically long ellipsoidal feature centered at the \pntX{Z}{\pm 2,0,44} points, but the intensity of the feature is very weak at the \pntX{Z}{0,0,42} point.
Furthermore, there are enhanced intensities in the vicinity of the \pntX{\Gamma}{\pm 2, 0, 42} points, but there is no corresponding feature in the vicinity of \pntX{\Gamma}{0,0,44} point.
They are considered as the contributions from PSF effect as already shown in the band structures shown in Fig.~\ref{ARPES_hn_band}.

To avoid the influence from this effect, we have symmetrized and folded the Fermi surface map within the momentum region of $k_x =-2$ -- $2 (\pi/a)$ and $k_z =42$ -- $44 (\pi/c)$.
The left panel of Fig.~\ref{ARPES_hn_FS} (b) shows the symmetrized and folded Fermi surface map.
The shapes of some Fermi surfaces are well recognized in this intensity map.
We have further analyzed this Fermi surface map by using the 2D-curvature method \cite{curvature}, and the result is shown in the right panel of Fig.~\ref{ARPES_hn_FS} (b).
The locations of $\alpha$, $\beta$, and $\gamma$ bands derived from the band structures shown in Fig.~\ref{ARPES_hn_band} (a) and (c) are also indicated by filled circles.
The approximate shapes of the Fermi surfaces deduced from the Fermi surface map and the 2D-curvature analysis are indicated by dashed curves.
According to these images, there are one large Fermi surface centered at the \Gm point designated as $\alpha$ and two Fermi surfaces centered at the \pnt{Z} point designated as $\beta$ and $\gamma$.

We compare these experimental Fermi surfaces with the result of the band structure calculation.
Figure~\ref{ARPES_hn_FS} (c) shows the Fermi surface obtained by the band structure calculation and the simulated Fermi surface map based on the band structure calculation.
Within this high-symmetry plane, there exist four Fermi surfaces formed by bands 15--18.
Bands 15 and 16 form small football-shaped Fermi surfaces centered at the \pnt{Z} point.
Band 17 appears as the cross-section of the ``pillow-shaped'' Fermi surface, and it has a horizontally long ellipsoidal shape.
Band 18 appears as an electron Fermi surface with a horizontally-long ellipsoidal shape, which corresponds to the horizontal joint of the ``jungle-gym'' Fermi surface.
A comparison between the experimental Fermi surface map and the calculation result suggests that there are certain correspondences between them.
For example, the shape of the horizontally long large Fermi surfaces $\alpha$ and $\beta$ in the experimental Fermi surface map correspond well to the Fermi surfaces of bands 18 and 17, respectively.
Furthermore, the vertically long ellipsoid feature around the \pntX{Z}{2,0,44}, designated as $\gamma$, also corresponds to the calculated Fermi surfaces 15 and 16.
On other other hand, there are some disagreements between the experimental and the calculated Fermi surface map.
For example, there is an enhanced intensity around the \Gm point in the experimental Fermi surface maps although there is no corresponding features in the calculation.
This might be due to the contribution from a surface state as discussed in Fig.~\ref{ARPES_hn_band} (a), but its origin is not clear.

Accordingly, the band structures in the vicinity of the Fermi energy and the topologies of Fermi surfaces within the $k_x$ -- $k_z$ plane are qualitatively described by the band structure calculation.
The topologies of the football-shaped Fermi surfaces around the \pnt{Z} point and the large Fermi surfaces formed by bands 17 and 18 around the \pnt{Z} and the \Gm points agrees with those obtained by band structure calculation although the number of small hole pocket in the very vicinity of the \pnt{Z} point is not well resolved in the present spectra.

\begin{figure}
	\includegraphics[scale=0.5]{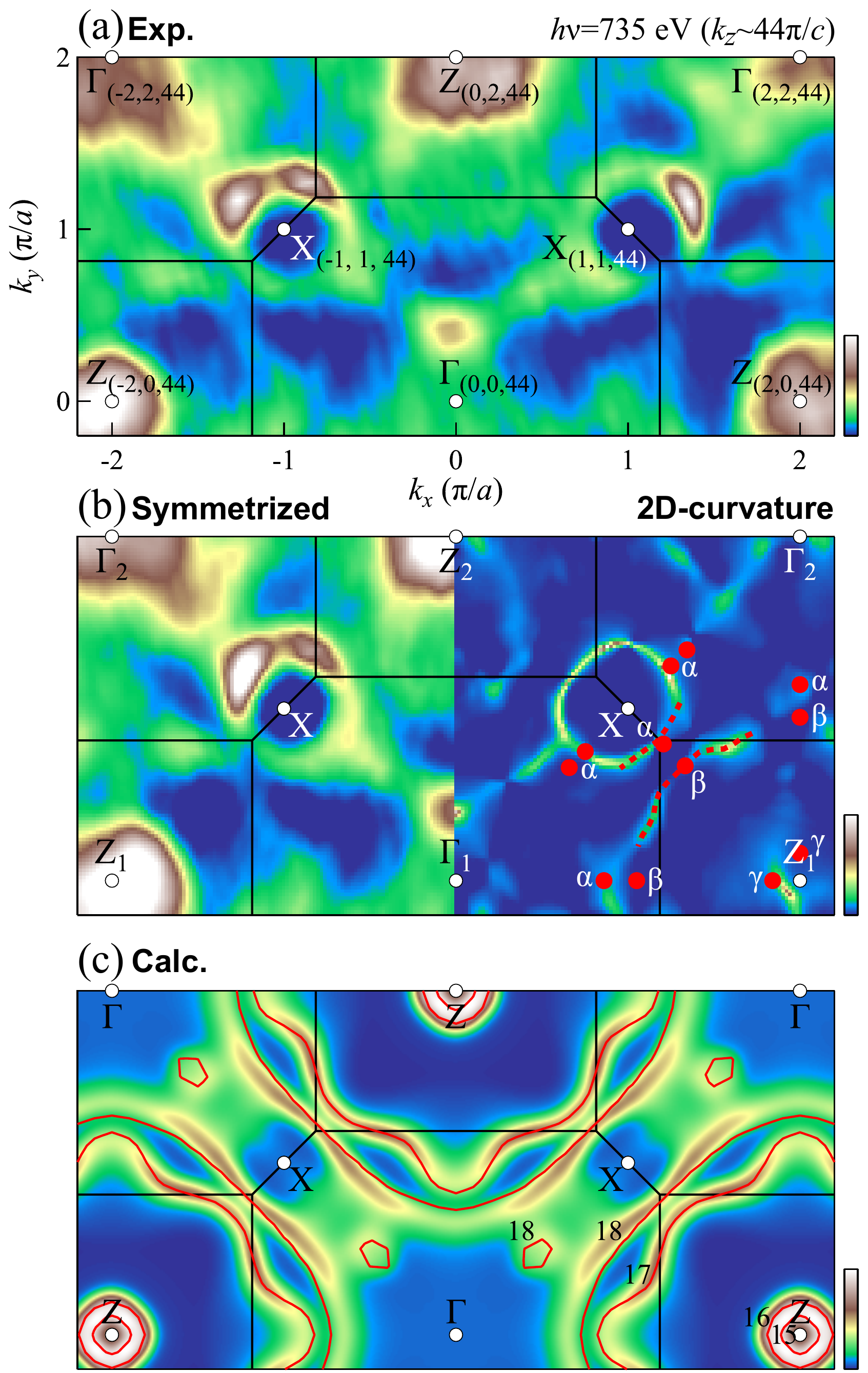}
	\caption{(Online color)
	Fermi surfaces of \ThRuSi obtained by angular scan of ARPES spectra measured at \hn{=735}, together with the corresponding results of the band structure calculation.
	(a) Fermi surfaces of \ThRuSi obtained by integrating ARPES spectra over $\mathrm{100~meV}$ around \EF.
	(b) Fermi surfaces of \ThRuSi symmetrized relative to $k_x=0$ (left panel) and Fermi surface obtained by the 2D curvature method \cite{curvature} (right panel).
	The Fermi momenta deduced from ARPES spectra shown in Figs.~\ref{ARPES_hn_band} and Figs.~\ref{ARPES_kxky_band} are plotted by filled circles.
	Dashed curves are guide to eyes.
	(c) Calculated Fermi surfaces and the simulated Fermi surfaces within the $k_x-k_y$ plane.
	}
\label{ARPES_kxky_FS}
\end{figure}
\subsection{Electronic structure within $k_x$--$k_y$ plane}
To further understand the shapes of the Fermi surfaces, we have measured ARPES spectra within the $k_x$--$k_y$ plane.
First, we focus on the electronic structure around the \pnt{X} point.
Figure~\ref{ARPES_kxky_band} (a) shows ARPES spectra of \ThRuSi measured along the \lineXXX{\Gamma}{X}{\Gamma} high-symmetry line.
These spectra were measured at \hn{= 735}, which correspond to the ARPES cut with $k_z \sim 44 (\pi / a)$.
There is a parabolic energy dispersion centered at the \pnt{X} point, whose approximate location is indicated by the dashed curves in the figure as band $\alpha$.
It forms a small hole pocket in between the \pnt{X} and the \Gm points.
Figure~\ref{ARPES_kxky_band} (b) displays the calculated band dispersion and the simulated ARPES spectra.
In the calculation, band 18 also forms a small electron pocket in between the \pnt{X} and the \Gm points, and its behavior is very similar to the experimentally observed band $\alpha$, although its momentum position and band width are somewhat different.

Figure~\ref{ARPES_kxky_band} (c) shows the experimental ARPES spectra measured along the \lineXXX{Z}{X}{Z} high-symmetry line, which is perpendicular to the \lineXXX{\Gamma}{X}{\Gamma} high-symmetry line.
The momentum distribution curve (MDC) at \EF is also shown in the upper panel.
The structure around the \pnt{X} point is very similar to that of \lineXXX{\Gamma}{X}{\Gamma} high-symmetry line, but there exist two bands $\alpha$ and $\beta$, that cross the Fermi energy.
The existence of the bands $\alpha$ and $\beta$ can be understood from the MDC spectrum at \EF where two peaks exist in this region.
The band $\beta$ is missing in the left half of the ARPES spectra probably due to the PSF effects.
The band structure calculation shown in Fig.~\ref{ARPES_kxky_band} (d) also predicts two electron-like bands 17 and 18 centered at the \pnt{X} point which form Fermi surfaces.
Their structures in the vicinity of \EF are very similar to the experimental bands $\beta$ and $\alpha$ in Fig.~\ref{ARPES_kxky_band} (c) although their band widths are different.
Therefore, although there are some quantitative differences, the essential topology of Fermi surfaces around the \pnt{X} point agrees well with the band structure calculation.

Next, we show the Fermi surface map within the $k_x$ -- $k_y$ high symmetry plane to further understand the overall in-plane shape of the Fermi surfaces.
Figure~\ref{ARPES_kxky_FS} (a) shows the Fermi surface map measured along the $k_x$--$k_y$ plane obtained by integrating at \EF of the ARPES spectra measured at \hn{=735} over $100~\mathrm{meV}$.
There are a few characteristic features in the Fermi surface map.
The most prominent one is the circular-shaped enhanced intensity centered at the \pnt{X} point, which was also observed in the previous ARPES study with $h\nu=17$ and $30~\mathrm{eV}$ \cite{Denlinger_XRu2Si2}.
There is also an enhanced intensity in the vicinity of the \pnt{Z} point, which corresponds to the hole-pocket-like features $\gamma$ observed in the ARPES spectra shown in Fig.~\ref{ARPES_hn_band} (a). 
There exist complex features around the \Gm point, and its appearance is considerably different between the \pntX{\Gamma}{0,0,44} and the \pntX{\Gamma}{\pm 2,2,44} points.

\begin{figure*}[ht]
	\includegraphics[scale=0.5]{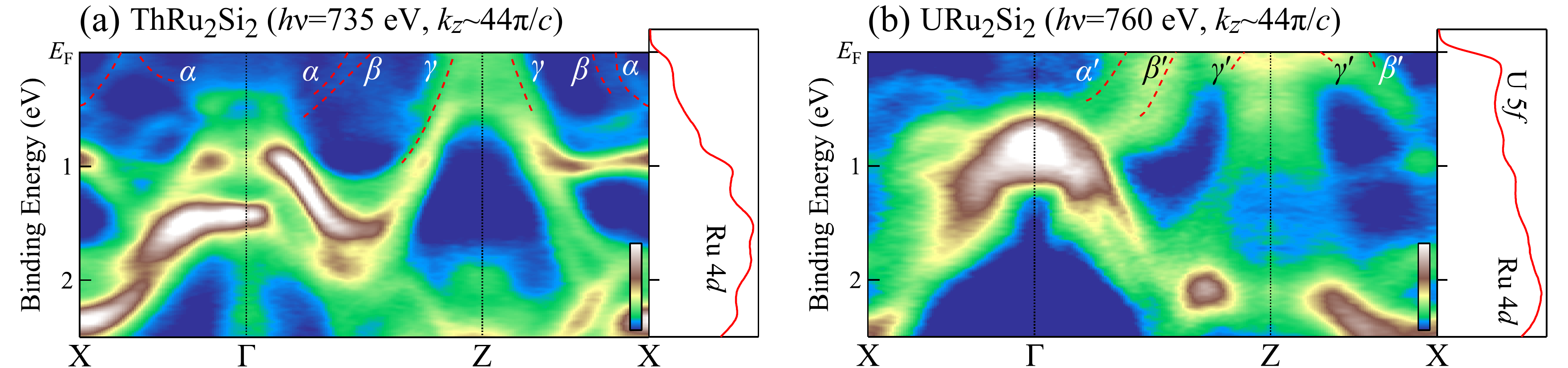}
	\caption{(Online color)
	Comparison of ARPES spectra of \ThRuSi and \URuSi
	The spectra are divided by the Fermi--Dirac function broadened by instrumental energy resolution to avoid the influences of Fermi cut-off.
	The AIPES spectra obtained by integrating the ARPES spectra within $k_x = -2$ -- $2 (\pi/a)$ and $k_y = 0$ -- $2 (\pi/a)$ are shown in the right panel of each figure.
	(a) ARPES spectra of \ThRuSi measured along the $\mathrm{X}$ -- $\mathrm{\Gamma}$ -- $\mathrm{Z}$ -- $\mathrm{X}$ high-symmetry line at \hn{=735}.
	(b) ARPES spectra of \URuSi measured along the $\mathrm{X}$ -- $\mathrm{\Gamma}$ -- $\mathrm{Z}$ -- $\mathrm{X}$ high-symmetry line at \hn{=760}.
	Data replotted from Ref.~\cite{SF_review_JPSJ,URu2Si2_SXARPES}, and bands $\alpha'$, $\beta'$, and $\gamma'$ in this figure correspond to band C, B (and E), and A (and D) in Fig.~2 of Ref.~\cite{URu2Si2_SXARPES}, respectively.
	Note that the bands $\alpha'$, $\beta'$, and $\gamma'$ should be renormalized in the very vicinity of \EF, but their behaviors are not well resolved due to the instrumental energy resolution.
	}
\label{ThRu2Si2_URu2Si2}
\end{figure*}
To further reveal the in-plane Fermi surface topology, we have symmetrized the Fermi surface map relative to $k_x=0$, and made its 2D-curvature analysis as shown in the left and right panels of Fig.~\ref{ARPES_kxky_FS} (b).
The Fermi momenta estimated from ARPES spectra measured along the \lineXXX{\Gamma}{Z}{\Gamma} (Figs.~\ref{ARPES_hn_band} (c)), the \lineXXX{\Gamma}{X}{\Gamma} (Fig.~\ref{ARPES_kxky_band} (a)), and the \lineXXX{Z}{X}{Z} (Fig.~\ref{ARPES_kxky_band} (c)) high-symmetry lines are also plotted by filled circles.
In the result of the 2D-curvature analysis, there is an arc-shaped feature in between the $\mathrm{Z_1}$ and the $\mathrm{X}$ points, which corresponds to the part of a hole-type Fermi surface formed by band $\beta$ centered at the \pnt{Z} point.
Although the feature around the \pnt{X} point has a circular shape, the behaviors of the bands $\alpha$ forming this feature are different between the \lineXX{X}{\Gamma} and the \lineXX{X}{Z} directions; it forms a small hole-pocket along the \lineXX{X}{\Gamma} direction while two bands $\alpha$ and $\beta$ form electron-type Fermi surfaces around the \pnt{X} point as shown in Fig.~\ref{ARPES_kxky_band}(a) and (c).
Furthermore, band $\alpha$ forms Fermi surface along the \lineXX{\Gamma}{Z} direction, and how they are connected within the plane was not well understood since its intensity is missing in many part of the Brillouin zone.
There are complex features around the $\mathrm{\Gamma_1}$ and the $\mathrm{\Gamma_2}$ points, but their appearances are considerably different each other.
In the symmetrized map, there are enhanced intensities around $\Gamma_2$ point, but there is no corresponding feature around the $\Gamma_1$ point.
The 2D-curvature map also exhibit complex features around these points, but we could not understand the nature of Fermi surface.
One possibility is that they might be the contributions from surface states as finite intensities were observed around the \Gm point shown in Fig.~\ref{ARPES_hn_band} (a).

The topology of the Fermi surface was not well resolved from the experimental Fermi surface map, and we consider it with the comparison with the result of the band structure calculation.
Figure~\ref{ARPES_kxky_FS} (c) shows the Fermi surface and the simulated Fermi surface map based on the band structure calculation.
There exist two concentric hole pockets formed by bands 15 and 16, and the cross-section of the ``pillow'' formed by band 17 centered at the \pnt{Z} point.
Band 18 forms a large open Fermi surface and a small hole-pocket between the \Gm and the \pnt{X} points.
A comparison between the experimental results and the calculation suggests that there are similar features.
For example, there are enhanced intensities around the \pnt{Z} point in both the experimental and the calculated Fermi surface map, which originate from the hole-like Fermi surface around the \pnt{Z} point.
In addition, the Fermi surface formed by band $\beta$ has a very similar topology to the calculated Fermi surface formed by band 17.
The Fermi surfaces formed by band $\alpha$ have some common features with the calculated Fermi surface formed by band 18; it forms a small hole-pocket along the \lineXX{\Gamma}{X} high symmetry line and large hole-like Fermi surface centered at \pnt{Z} point.
On the other hand, the intensity of band $\alpha$ is missing in many part of the Brillouin zone, and an overall agreement between experimental band $\alpha$ and calculated band 18 was not well understood.
Accordingly, there are some similar features in the experimental and calculated Fermi surface maps and band structure although there are some unexplained features particularly around \Gm point.

\subsection{Discussion and comparison with \URuSi}
We measured the ARPES spectra of \ThRuSi within the $k_x$--$k_z$ and $k_x$--$k_y$ planes, and the topologies of the Fermi surfaces were mostly explained by the band structure calculation although the in-plane topology of the Fermi surface formed by band $\alpha$ was not clear.
Here we discuss the relationship between the present result and dHvA studies.
In general, the locations of Fermi surface in Brilloin zone cannot be determine by the dHvA, but the size of Fermi surface determined by the ARPES experiment is not as accurate as the one obtained by dHvA experiment owing to various contributions such as instrumental energy resolution, momentum resolution, and the broadening along $k_\perp$ direction \cite{Strocov}.
In the dHvA studies by Matsumoto \etal \cite{ThRu2Si2_dHvA2}, branches from three Fermi surfaces $\alpha$, $\beta$, and $\gamma$ were observed, which correspond to the Fermi surfaces $\alpha$, $\beta$, and $\gamma$ observed in the present ARPES measurement, respectively.
Thus, the result of the present ARPES study is consistent with the dHvA studies concerning the number of Fermi surfaces.
Furthermore, the locations of the Fermi surfaces formed by bands $\beta$ and $\gamma$ in the Brillouin zone were determined by the present study, and they were consistent with the result of the band structure calculation.
Although the topology of the in-plane Fermi surface formed by band $\alpha$ was not completely determined from the present ARPES spectra, the branch $\alpha$ is in the dHvA study was well explained by the band structure calculation, suggesting the topology of Fermi surface $\alpha$ also can be explained by the band structure calculation.
Therefore, by combining the results of the ARPES and dHvA studies, it is concluded that the topology of the Fermi surface of \ThRuSi is mostly explained by the band structure calculation.

Finally, we compare the present result with the ARPES study of \URuSi.
Figures~\ref{ThRu2Si2_URu2Si2} (a) and (b) show the ARPES spectra of \ThRuSi and \URuSi measured along the $\mathrm{X}$ -- $\mathrm{\Gamma}$ -- $\mathrm{Z}$ -- $\mathrm{X}$ high-symmetry line.
The photon energies used were \hn{=735} for \ThRuSi and \hn{=760} for \URuSi, which correspond to the ARPES cut with $k_z \sim 44 (\pi/c)$.
Therefore, the inequalities originating from PSF effects between the ARPES spectra of \ThRuSi and \URuSi are absent in the comparison although the slightly different photoionization cross section might be expected.
The spectra of \URuSi were replotted from Ref.~\cite{SF_review_JPSJ,URu2Si2_SXARPES}, and bands $\alpha'$, $\beta'$, and $\gamma'$ in this figure correspond to band C, B (and E), and A (and D) in Fig.~2 of Ref.~\cite{URu2Si2_SXARPES}, respectively.
The AIPES spectra obtained by integrating the ARPES spectra within $k_x = -2$ -- $2 (\pi/a)$ and $k_y = 0$ -- $2 (\pi/a)$ are shown in the right panel of each figure.
The ARPES spectra are divided by the Fermi--Dirac function broadened by instrumental energy resolution to avoid the influences of Fermi cut-off.
Sample temperature was $T=20~\mathrm{K}$, and both samples were in the paramagnetic phase.

The ARPES spectra of \ThRuSi and \URuSi have very similar structures, but there are many notable differences in their details. 
On the high-binding-energy side, the bands with large contributions from the \orb{Ru}{4d} states in the binding energy of \EB{=0.5-2} have similar cosine-shaped dispersions, but their details differ considerably.
For example, the dispersion has enhanced intensity around the \Gm point at \EB{\sim 1.5} in \ThRuSi, while it is weak in \URuSi and the state at \EB{\sim 0.9} is enhanced instead.
The structure of the band around the \pnt{Z} point is also different between them.
These suggest that the \Uf states are considerably hybridized with ligand states, and they substantially influence the entire band structure.
This is in strong contrast with the case of \CeRuSi and \LaRuSi \cite{Denlinger_XRu2Si2}, where most of $\mathrm{Ce}~4f$ states are distributed on the high-binding-energy side as localized $f^0$ peak, and ARPES spectra are essentially identical except for the contribution of heavy quasi-particle bands located within \EB{\lesssim 0.2}.
This suggests that \Uf states are considerably hybridized with the ligand states in \URuSi while \Cef states are less hybridized in \CeRuSi.

More importantly, the structures in the vicinity of \EF are very different between \ThRuSi and \URuSi.
The spectra of \URuSi have more complicated structures than those of \ThRuSi.
One of the most prominent differences between these two sets of spectra is the dispersive features along the \lineXX{\Gamma}{Z} high-symmetry line.
The intensity of bands $\alpha$ and $\beta$ is very weak in the spectra of \ThRuSi, while the intensity of the corresponding bands $\alpha'$ and $\beta'$ is significantly enhanced in the spectra of \URuSi.
This suggests that bands $\alpha'$ and $\beta'$ have large contributions from \Uf states, indicating the itinerant nature of \Uf states in \URuSi.
Moreover, the spectral profiles in the very vicinity of \EF at the \pnt{Z} point are also different between \ThRuSi and \URuSi.
Band $\gamma$ has relatively simple hole-like dispersion in \ThRuSi, while the corresponding band $\gamma'$ exhibits more complex behavior just below \EF in \URuSi, suggesting that the hybridization is anisotropic in the momentum space.
The strongly hybridized nature is consistent with the result of previous ARPES studies \cite{URu2Si2_Tomasz_review, SF_review_JPCM,SF_review_JPSJ}.

\section{CONCLUSION}
The electronic structure of \ThRuSi was studied by AIPES and ARPES.
The $\mathrm{Th}~4f$ core-level spectrum is very similar to that of $\mathrm{Th}$ metal, suggesting that \Th is tetravalent in \ThRuSi.
The three-dimensional shapes of the Fermi surfaces and the band structure were obtained experimentally by ARPES.
Two hole-like Fermi surfaces around the \pnt{Z} point and one electron-like Fermi surface around the \Gm point were observed, and the results are mostly consistent with the result of the dHvA study. 
The topology of the Fermi surface and the band structure of \ThRuSi were mostly explained by the band structure calculation.
Meanwhile, the structure of the hole-pocket Fermi surface around the \pnt{Z} point was not well resolved, and the number of Fermi surfaces was not determined.
Comparison of AIPES and ARPES spectra between \ThRuSi and \URuSi showed that they have very different natures especially in the vicinity of \EF.
This indicates that the \Uf electrons in \URuSi are strongly hybridized with ligand states, suggesting that \Uf electrons in \URuSi have a considerably itinerant nature.

\acknowledgments
The experiments were performed under Proposal Nos. 2015A3820, 2015B3810, 2016A3820 and 2016B3811 at SPring-8 BL23SU.
The present work was financially supported by JSPS KAKENHI Grant Numbers 21740271, 26400374, 16H01084 (J-Physics), and 16K05463; and Grants-in-Aid for Scientific Research on Innovative Areas ``Heavy Electrons'' (Nos. 20102002 and 20102003) from the Ministry of Education, Culture, Sports, Science, and Technology, Japan.

\bibliographystyle{apsrev}
\bibliography{ThRu2Si2}

\end{document}